\documentclass[10pt,letterpaper,twocolumn]{article}
\usepackage{ol2}
\usepackage{amsmath}

\renewcommand{\Im}{\mathrm{Im}}
\renewcommand{\Re}{\mathrm{Re}}
\renewcommand{\d}{\mathrm{d}}
\newcommand{\e}{\mathrm{e}}

\begin{document}
\twocolumn[%
\title{Generation and amplification of Raman Stokes and anti-Stokes waves}

\author{E. Brainis,$^{1,*}$ S. Clemmen,$^{2}$ and Serge Massar$^{2}$}
\address{$^1$Optique et acoustique, CP 194/5, Universit\'e libre de Bruxelles,%
avenue F.D. Roosevelt 50,  1050 Brussels, Belgium}%
\address{$^2$Laboratoire d'information
quantique, {CP} 225, Universit\'{e} libre de Bruxelles, boulevard du
Triomphe, 1050 Brussels, Belgium}
\address{$^*$Corresponding author: ebrainis@ulb.ac.be}

\begin{abstract}
We present general analytical expressions of Stokes and anti-Stokes
spectral photon-flux densities that are spontaneously generated by a
single monochromatic pump wave propagating in a single-mode optical
fiber. We validate our results by comparing them with experimental
data. Limiting cases of the general expressions corresponding to
interesting physical situations are discussed.
\end{abstract}
 \maketitle
  ]

 The third
order nonlinearity seen by light propagating in optical fibers is
composed of two parts: a quasi instantaneous contribution coming
from the electronic response, and a delayed Raman response coming
from the coupling to molecular vibrations (phonons). The influence
of these two terms on the propagation of light are often considered
separately, see for instance the treatment in \cite{Agrawal}. But in
many cases they must be treated collectively, together with the
influence of dispersion which plays an essential role, particularly
when the phase matching conditions are (even approximately)
satisfied. That a holistic approach is necessary was first
understood by Bloembergen and Shen\cite{BS}, and confirmed
experimentally in later work. For instance it was shown that the
gain of the Raman Stokes wave is strongly affected by the
dispersion, and can even vanish when there is zero phase mismatch
\cite{GMPD,VEC}. In addition it was shown in \cite{BS}, and
confirmed experimentally in \cite{CWH}, that the Raman anti-Stokes
wave, which in the naive approach is exponentially damped, in fact
grows exponentially at late times. This effect has important
implications for supercontinuum generation in photonic crystal
fibers using long  pump pulses (ps and longer). In these systems
spontaneous amplification of vacuum fluctuations at the anti-Stokes
wavelengths can give rise to a blue detuned supercontinuum
\cite{CCLHKWR}, see \cite{rulkov,travers} for further studies and
\cite{DGC} for a review of supercontinuum generation.

A quantum description of light propagation in optical fibers,
including both the instantaneous Kerr non linearity and the Raman
scattering has been developed \cite{BKH} and used to quantify noise
limits on squeezing \cite{SB}, noise in $\chi^{(3)}$ parametric
amplifiers \cite{VK}, noise in coherent  anti-Stokes Raman
scattering \cite{DSCJ}, and noise in photon pair generation
experiments \cite{LYA,LYA2}. In the present work we apply this
quantum theory, using the formulation of \cite{VK}, to describe the
spontaneous growth of Raman Stokes and anti-Stokes waves from vacuum
fluctuations. Our analytical predictions are compared to
experimental results reported in \cite{CWH,CCLHKWR} which had
previously been fitted with semi-empirical formulae,  or numerical
calculations.

We consider a continuous pump with complex amplitude $\sqrt{P} \
\e^{i\phi(x,t)}$, where $P$ is the power flowing through the fiber
and $\phi(x,t)=-\omega_p t+[k_p+  \gamma P] x$. In this expression,
$\omega_p$ is the pump frequency and $k_p$ is the (linear) wave
number. Self-phase modulation gives a nonlinear contribution $\gamma
P$ to the total wave number. Here $\gamma$ is the third-order
nonlinearity parameter of the fiber \cite{Agrawal}. The quantum
field operator
\begin{equation}
\begin{split}
 A(x,t)&=
  \e^{i\phi(x,t)}
\int_0^\infty \left(\sqrt{\frac{\hbar \omega_s}{2 \pi}} \
A_s(\Omega,x) \  \e^{i\Omega t} \right.\\
&\phantom{=}+ \left.\sqrt{\frac{\hbar \omega_a}{2 \pi}} \
A_a(\Omega,x) \ \e^{-i\Omega t}\right)  \d\Omega + \mathrm{h.c.}
\label{field}
\end{split}
\end{equation}
describes the perturbations around this stationary solution through
the combined effect of Raman scattering and four-wave mixing. These
are composed of symmetrically detuned Stokes and anti-Stokes photons
with respective frequencies $\omega_s = \omega_p-\Omega$ and
$\omega_a = \omega_p+\Omega$, and wave numbers $k_s$ and $k_a$. The
operators $A_s(\Omega,x)$ and $A_a(\Omega,x)$ are \emph{destruction
operators} for Stokes and anti-Stokes photons. Their equations of
motion can be deduced from the quantum theory of light propagation
and solved analytically (so long as the perturbation field is small
compared to the pump)\cite{VK}.

Once $A_s(\Omega,x)$ and $A_a(\Omega,x)$ are known, any physical
quantity can be computed. Here we are concerned by the mean
\emph{spectral photon-flux densities}
%\begin{equation*}
%\frac{1}{2\pi
%\epsilon}\int^{\Omega+\frac{\epsilon}{2}}_{\Omega-\frac{\epsilon}{2}}\int^{\Omega+\frac{\epsilon}{2}}_{\Omega-\frac{\epsilon}{2}}
%\langle A_{s,a}^\dag(\Omega_1,x)A_{s,a}(\Omega_2,x)\rangle
%\d\Omega_1 \d\Omega_2
%\end{equation*}
%for $\epsilon \to 0$.
\begin{equation}
\begin{split}
 &f_{s,a}(\Omega,x)=\lim_{\epsilon\rightarrow 0} \ \frac{1}{2\pi
\epsilon}\\
 &\times \int^{\Omega+\frac{\epsilon}{2}}_{\Omega-\frac{\epsilon}{2}}\int^{\Omega+\frac{\epsilon}{2}}_{\Omega-\frac{\epsilon}{2}}
\langle A_{s,a}^\dag(\Omega_1,x)A_{s,a}(\Omega_2,x)\rangle
\d\Omega_1 \d\Omega_2.
\end{split}
\end{equation}
Using the approach of \cite{VK}, we find
\begin{subequations}\label{f}\allowdisplaybreaks
\begin{align}
\begin{split}
f_s(\Omega,x)  &= \frac{1}{2\pi} \ \frac{|\chi(\Omega)|^2}{|\kappa(\Omega)|^2} \ |\sinh(\kappa(\Omega) x)|^2\\
&\phantom{=}+ \frac{|\Im[\chi(\Omega)]|}{\pi} \  \rho_+(\Omega,x) \
\left(n(\Omega)\negmedspace+\negmedspace 1\right)
\end{split} \\
\begin{split}
f_a(\Omega,x)  &= \frac{1}{2\pi} \ \frac{|\chi(\Omega)|^2}{|\kappa(\Omega)|^2} \ |\sinh(\kappa(\Omega) x)|^2\\
&\phantom{=}+ \frac{|\Im[\chi(\Omega)]|}{\pi} \  \rho_-(\Omega,x) \
n(\Omega),
\end{split}
\end{align}
\end{subequations}
with
\begin{equation}
\rho_{\pm}(\Omega,x)=\int_0^x \d x' \left|\cosh(\kappa(\Omega)
x')\pm  i \frac{\Delta k(\Omega)}{2\kappa(\Omega)} \sinh(\kappa
x')\right|^2.\label{rhoa}
\end{equation}
The photon fluxes depend on three basic parameters: the pump power
$P$, the detuning $\Omega$ and the fiber temperature $T$. The pump
power and the nonlinear response of the fiber are combined as
$\chi(\Omega)= \gamma P \left[ (1-f_r)+f_r\chi_r(\Omega)\right]$,
where the nonlinearity is decomposed into an instantaneous
(electronic) part and a retarded (molecular or Raman) one.
$f_r\approx 0.18$ is the fraction of the total nonlinearity due to
the Raman scattering and $\chi_r(\Omega)$ is the normalized spectral
Raman response ($\Re[\chi_r(0)]=1$, $\Im[\chi_r]<0$). The complex
parameter $\kappa(\Omega)=[-\left(\Delta k(\Omega)/2\right)^2-\Delta
k(\Omega) \chi(\Omega)]^{1/2}$ controls the growth rate of the
Stokes and anti-Stokes waves. It depends both on $\chi(\Omega)$ and
on the linear phase-mismatch $\Delta k(\Omega)=k_s + k_a - 2 k_p$.
When the fiber is pumped sufficiently far away from the zero
dispersion wavelength, the phase-mismatch $\Delta k(\Omega)$ is well
approximated by $\beta_2\Omega^2$, where $\beta_2$ is the
group-velocity dispersion parameter. The real part of
$\kappa(\Omega)$ is chosen positive so that it can be interpreted as
the \emph{amplification gain} for small signals at frequencies
$\omega_p\pm\Omega$. Finally, the temperature $T$ influences the
photon fluxes through the factor
\begin{equation}\label{n}
n(\Omega) = \left(\exp(\hbar \Omega / k_B T)-1\right)^{-1}
\end{equation}
related to the \emph{phonon population}.

The integral in Eq.~(\ref{rhoa}) can be carried out exactly,
leading to somewhat cumbersome expressions. Here however, we prefer
to focus on two physically important limits : the \emph{spontaneous
scattering limit} ($|\kappa| x\rightarrow 0$) and the
\emph{stimulated amplification and wave-mixing limit} ($\Re[\kappa]
x\rightarrow \infty$). The first one is of great importance for
photon-pair generation, while the second one applies to
supercontinuum generation.

Considering the limit of small $|\kappa| x$, Eqs.~(\ref{f}) give
\begin{subequations}\label{spRS}
\begin{eqnarray}
f_s &\approx&\frac{1}{\pi}|\Im[\chi]| \ x \ (n+1) \\
f_a &\approx&\frac{1}{\pi}|\Im[\chi]| \ x \ n,
\end{eqnarray}
\end{subequations}
up to the first order in $|\kappa| x$. This contribution to the
total photon flux is called the \emph{spontaneous Raman scattering}.
Stokes and anti-Stokes photons are emitted independently. A Stokes
photon emission is always accompanied by the emission of a phonon of
energy $\hbar\Omega$, in contrast to the anti-Stokes process for
which conservation of energy requires a phonon to be absorbed. For
this reason, the anti-Stokes process in inhibited when the phonon
population is zero (low temperature limit). Using Eq.~(\ref{n}), one
finds that $\lim_{|\kappa|x \to 0}f_a/f_s=\exp[-\hbar\Omega/(k_BT)]$
in accordance with the usual formulation theory of spontaneous Raman
scattering in fibers (see \cite{CWH} and references within). Upon
adding the extra term $|\chi|^2x^2/(2\pi)$ to Eqs.~(\ref{spRS}a) and
(\ref{spRS}b) one accounts for the \emph{spontaneous four-photon
scattering} process in which a Stokes photon and an anti-Stokes
photon are emitted together after the absorbtion of two pump
photons. This is the process used for photon-pair generation in
fibers. Since it is only of second order in $|\kappa| x$,
spontaneous Raman scattering always plays a detrimental role in
photon-pair generation experiments and is referred to ``Raman
noise'' (see \cite{LYA,LYA2} for a more thorough discussion). It is
interesting to note that linear dispersion has no impact on the
values of the photon fluxes up to the second order in $|\kappa|x$
since they do not depend on $\Delta k$. However, the validity of the
Maclaurin expansion is restricted to small values of $\Delta k$
because $|\kappa|\rightarrow (\Delta k/2)$ for fixed pump power and
large $\Delta k$ values.

The asymptotic behavior for $\Re[\kappa]x\rightarrow \infty$ is also
simple since we keep only the exponentially growing terms in
Eqs.~(\ref{f}) :
\begin{subequations}\label{fsfa}
\begin{align}
%\begin{split}
f_s &\sim \frac{ \e^{2\Re[\kappa]x}}{8 \pi} \left(
\frac{|\chi|^2}{|\kappa|^2}\negthinspace+\negthinspace
\frac{|\Im[\chi]|}{\Re[\kappa]}\frac{|\kappa + i \frac{\Delta
k}{2}|^2}{|\kappa|^2}
 (n+1) \right)\\
f_a &\sim \frac{ \e^{2\Re[\kappa]x} }{8 \pi} \left(
\frac{|\chi|^2}{|\kappa|^2}\negthinspace+\negthinspace
\frac{|\Im[\chi]|}{\Re[\kappa]}\frac{|\kappa - i \frac{\Delta
k}{2}|^2}{|\kappa|^2}
 \ n \ \right).
%\end{split}
\end{align}
\end{subequations}
Note that $|\kappa + i \frac{\Delta k}{2}| > |\kappa - i
\frac{\Delta k}{2}|$ which implies that the Stokes wave is always
stronger than the anti-Stokes wave, as expected.
\begin{figure}[b]
\centerline{\includegraphics[width=8.4cm]{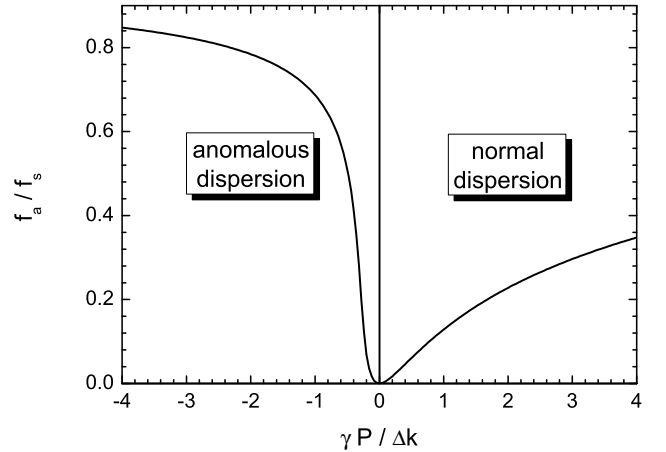}}
\caption{Ratio of anti-Stokes to Stokes photon fluxes, at the peak
of the Raman gain, as a function of $\gamma P / \Delta k$.}
\label{fig1}
\end{figure}
In Fig.~\ref{fig1}
we have plotted the ratio $f_a / f_s$ of the anti-Stokes to Stokes
fluxes at the peak of the Raman gain ($\Omega/(2 \pi)=13.2$~THz) at
$T=300$~K as a function of $\gamma P/\Delta k$. In these
circumstances, $n=0.14$ and $\chi = \gamma P \ (0.82 - i \ 0.25)$
for silica optical fibers. When $\gamma P/|\Delta k| \gg 1$ pair
creation dominates over Raman scattering in Eqs.~(\ref{fsfa}), both
in the normal and anomalous dispersion regimes, and we have
\begin{equation}\label{limsup}
f_s \simeq f_a  \sim\frac{ \e^{2\Re[\kappa]x}}{8 \pi}
\frac{|\chi|^2}{|\kappa|^2}.
\end{equation}
However $f_a/f_s$ tends much faster to $1$ in the anomalous
dispersion regime than in the normal dispersion regime, as is
apparent from Fig.~\ref{fig1}. When the opposite limit $\gamma
P/|\Delta k| \ll 1$ is considered, the approximation $\kappa \simeq
|\Im[\chi]|+i (\Delta k/2 + \Re[\chi]) + o(|\chi/\Delta k|)$ holds
and
\begin{subequations}\label{liminf}
\begin{align}
f_s &\sim \frac{ \e^{2|\Im[\chi]|x}}{2 \pi} \ (n+1)\\
f_a &\sim \frac{ \e^{2|\Im[\chi]|x}}{2 \pi} \ \frac{|\chi|^2}{\Delta
k^2} \ (n+1).
\end{align}
\end{subequations}
The ratio of anti-Stokes to Stokes fluxes is then given by
${f_a}/{f_s}= {|\chi|^2}/{\Delta k^2}$. It is worth noting that it
is \emph{independent of temperature}. This simple ratio was first
exhibited in \cite{CWH}. However, it should be noted that its origin
is more complicated than the argument given in \cite{CWH} since the
different processes of pair creation and phonon emission/absorption
all contribute to it.

Even though Eqs.~(\ref{fsfa}), (\ref{limsup}) and (\ref{liminf}) are
valid for arbitrary detunings $\Omega$, we now focus on $\Omega/(2
\pi)=13.2$~THz (corresponding to the Raman peak) in order to compare
theory to experiments.

\begin{figure}[t]
\centerline{\includegraphics[width=8.4cm]{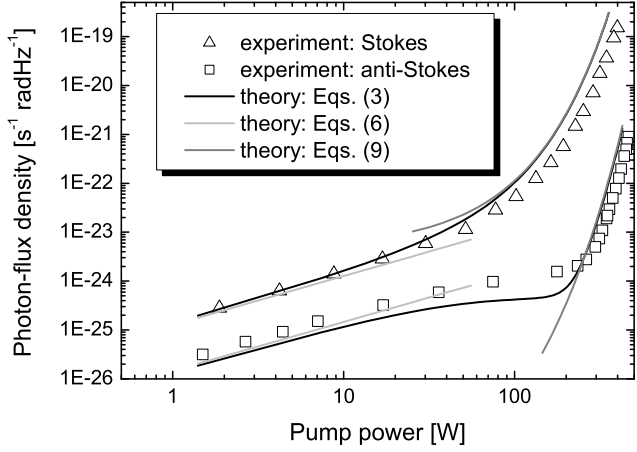}}
\caption{Evolution of the Stokes and anti-Stokes spectral
photon-flux densities with the pump power $P$. The parameters are:
$\gamma=23$~W$^{-1}$km$^{-1}$, $x=2.85$~m,
$\beta_2=50$~ps$^2$km$^{-1}$ and $T=293$~K. (Experimental data from
\cite{CWH}.)} \label{fig2}
\end{figure}
Experimental Stokes and anti-Stokes spectral photon-flux densities
in the $\gamma P/|\Delta k| \ll 1$ regime are available from
\cite{CWH}. They are plotted in Fig.~\ref{fig2} as a function of the
pump power $P$. For the pump powers that are considered $\gamma
P/|\Delta k| \le 0.032$. As seen from the figure, the asymptotic
formula (\ref{liminf}) [dark grey curves] apply for $P\ge300$~W
only. At lower pump power the condition $\Re[\kappa]x\gg 1$ is no
longer satisfied. For very low pump powers, one can however apply
the formula (\ref{spRS}) [light grey curves] since $|\kappa|x \ll
1$. Outside these limiting cases Eqs.~(\ref{f}) reproduce very well
the spectral photon-flux densities that are observed experimentally
[black curves].

In \cite{CCLHKWR}, experimental data are obtained for a photonic
crystal fiber with $\gamma = 150$~W$^{-1}$km$^{-1}$ and $\beta_2=7$
ps$^2$/km (normal dispersion). Measurements have been carried out
with $x=3$~m and $P=90$~W, as well as with $x=0.7$~m and $P=400$~W.
In both cases $\Re[\kappa]x\gg1$ but since $\gamma P/|\Delta k|$ is
close to one neither Eqs.~(\ref{limsup}) nor Eqs.~(\ref{liminf}) can
be used to compute the ratio $f_a/f_s$. Our result (\ref{fsfa})
should be used instead. It predicts the ratio $f_a/f_s=0.028$ for
$P=90$~W and $f_a/f_s=0.16$ for $P=400$~W ($T=300$~K). These are in
good agreement with the measured ratios ($f_a/f_s=0.016$ and $0.22$
respectively) given the uncertainty on the experimental parameters
in \cite{CCLHKWR} and probable polarization effects.

In summary we have developed an analytic theory to account for the
growth of the Stokes and anti-Stokes waves from the combined effects
of pair creation and Raman scattering. The results are in good
agreement with earlier experimental observations. They should find
applications in the optimization of supercontinuum sources based on
long pump pulses.

This research was supported by the EU project QAP (contract number
015848) and the IAP programme, Belgium Science Policy, under grant
P6-10. We thank S. Coen and D. A. Wardle for sharing with us their
experimental data.

%\end{multicols}

\end{document}